\documentclass[11pt,a4paper]{article}

\usepackage{epsfig,amsmath,amssymb,cite}
\usepackage{color}

\def\lp{\left(}
\def\rp{\right)}
\def\lb{\left[}
\def\rb{\right]}
\def\rk{\right\}}
\def\lk{\left\{}
\def\be{\begin{equation}}
\def\ee{\end{equation}}
\def\ba{\begin{array}}
\def\ea{\end{array}}
\def\bea{\begin{eqnarray}}
\def\eea{\end{eqnarray}}
\def\p{\partial}

\def\t{\theta}

\def\a{\alpha}
\def\b{\beta}
\def\g{\gamma}
\def\s{\sigma}

\def\l{\lambda}
\def\L{\Lambda}

\def\t{\tau}

\def\s{\sigma}
\def\S{\Sigma}
\def\e{\epsilon}
\def\v{\varphi}

\begin{document}

\title{Thin shells associated to black string spacetimes} 

\author{Ernesto F. Eiroa$^{1,2,}$\thanks{e-mail: eiroa@iafe.uba.ar}, Emilio Rub\'{\i}n de Celis $^{1,3,}$\thanks{e-mail: erdec@df.uba.ar},  
\begin{tiny}
{\normalsize •\begin{large}•\end{large}}\end{tiny}
Claudio Simeone$^{1,3,}$\thanks{e-mail: csimeone@df.uba.ar}\\
{\small $^1$ Departamento de F\'{\i}sica, Facultad de Ciencias Exactas y Naturales,} \\ 
{\small Universidad de Buenos Aires, Ciudad Universitaria Pabell\'on I, 1428, Buenos Aires, Argentina}\\
{\small $^2$ Instituto de Astronom\'{\i}a y F\'{\i}sica del Espacio (IAFE, CONICET-UBA),} \\ 
{\small Casilla de Correo 67, Sucursal 28, 1428, Buenos Aires, Argentina}\\
{\small $^3$ IFIBA--CONICET, Ciudad Universitaria Pabell\'on I, 1428, Buenos Aires, Argentina}} 

\maketitle

\begin{abstract}

In this article, we study thin shells of matter  connecting charged black string geometries with different values of the corresponding parameters. We analyze the matter content and the mechanical stability of the shells undergoing perturbations that preserve the cylindrical symmetry. Two different global configurations are considered: an interior geometry connected to an exterior one at the surface where the shell is placed, and two exterior geometries connected by a wormhole
throat located at the shell position.
\vspace{\baselineskip}


\end{abstract}

\section{Introduction}

Thin matter layers (or thin-shells) \cite{dar,israel} and their associated geometries appear in both cosmological and astrophysical frameworks. At a cosmological scale, the formalism used to define such layers has been applied in braneworld models, in which a spacetime is defined as the surface  where two higher dimensional manifolds are joined (see for instance \cite{davis} and references therein). At an astrophysical level, such matter layers appear, for example, as models for stellar atmospheres, gravastars, etc. \cite{brady,wilt}. Most thin-shell models considered in the literature are associated to spherically symmetric geometries. However, cylindrical shells and the corresponding spacetimes are also of physical interest.

Cosmic strings \cite{book} are topological defects which would result from symmetry breaking processes in the early Universe.  Theories with only the presence of scalar fields predict the so-called global strings, while the addition of one or more gauge fields leads to the prediction of local or gauge strings. The possible important role of cosmic strings in the explanation of structure formation at cosmological scale \cite{structure} and, besides, the fact that it would be possible to detect their gravitational lensing effects \cite{lens} led to a considerable amount of work devoted to their study \cite{strings}. A large proportion of the recent developments addressing cylindrical shells deals with geometries associated to cosmic strings \cite{ei1,ei2,eisi10,eoc12,eirdcsi,em16,cecemcl}. Other studies regarding cylindrical shells involve wormhole spacetimes, see for example   \cite{ri13,mahaa,eisi15,brokre}.

Black strings have a very different origin, as they would be associated to gravitational collapse. While for $3+1$ dimensions, in the absence of a cosmological constant, the collapse of a cylindrical matter distribution does not lead to horizons, within the framework of a theory with a negative cosmological constant, i.e. $\L<0$, the appearence of an event horizon does take place \cite{lemos1,lemos2}. In this sense, the situation is the same as in $2+1$ dimensional gravity, where an analogous of the Schwarzschild solution does not exist for zero cosmological constant, but for $\L<0$ we have the well known Ba\~nados--Teitelboim--Zanelli (BTZ) solution representing  a three-dimensional black hole \cite{BTZ}. In fact, the strong relation between these three and four-dimensional geometries was stressed in \cite{lemos3}, where it was shown that the BTZ black hole solution can be translated into the black string geometry. Different aspects of shells connecting  BTZ geometries were considered in \cite{eisi13,eisi14,lemos14,lemos15,btz15}.  Wormholes associated with charged black strings supported by shells with a Chaplygin equation of state were introduced in \cite{sharif}, while the linearized stability of thin-shell wormholes related to non charged black strings has been recently analyzed in \cite{black}. 

The four-dimensional action for gravity plus the  electromagnetic field is given by
\be S=\frac{1}{16\pi }\int d^4x\sqrt{-g}\left( R-2\L-F^{\mu\nu}F_{\mu\nu}\right),
\label{action}
\ee
where $g$ is the determinant of the metric, $R$ is the Ricci
scalar, and $F_{\mu\nu}=\partial_\mu A_\nu-\partial_\nu A_\mu$, with $A_\mu$ the electromagnetic vector potential. Under the assumption that the spacetime is static and cylindrically symmetric, the  corresponding field equations admit the solution (see Ref. \cite{lemos2})
\be
 ds^2 = - \left( \alpha^2r^2 - \frac{4m}{\alpha r}+\frac{4\lambda^2}{\alpha^2 r^2}\right) dt^2
+\left( \alpha^2r^2 - \frac{4m}{\alpha r}+\frac{4\lambda^2}{\alpha^2 r^2}\right)^{-1}dr^2
+r^2d\varphi^2
+\alpha^2 r^2 dz^2, \label{metric}
\ee
\be
A_\mu=\lp-\frac{2\l}{\alpha r}+\mathrm{const},0,0,0\rp,
\ee
where $m$ and  $\lambda$ are respectively the mass and charge per unit length, and $\alpha^2= -\Lambda /3 > 0$ is related to the cosmological constant. This geometry is singular at the axis of symmetry (the Kretschmann scalar diverges there). Depending on the values of these constants, we can have $g_{tt}=0$, so that the metric (\ref{metric}) can present an event horizon; thus the {\it black string} denomination. In particular, when $m=0$ and $\l =0$ no horizons are allowed, while if $m \neq 0$ and $\l =0$, there is only an event horizon located at $\a r_h = (4m)^{1/3}$. In case that $\l \neq 0$ the horizons correspond to the positive roots of the fourth degree polynomial $P(r) = \a ^4 r^4 - 4m \a r + 4 \l^2$; if $0<|\l | < \l _e = m^{2/3}\sqrt{3}/2$ there are two horizons satisfying $\a r_{-} < \a r_{+} < (4m)^{1/3}$, being $r_{+}$ the radius of the event one, while for the extremal charge per unit length $|\l | = \l _e$ these horizons fuse into one with radius $\a r_e = m^{1/3}$; finally if $|\l | > \l _e $ there are no horizons and only a naked singularity at the axis of symmetry is left. 

In the present work we address the study of shells associated to charged non-rotating black string geometries of the form (\ref{metric}). We will consider both the case of inner solutions joined to outer ones, and the case of traversable thin-shell wormholes connecting two exterior regions of manifolds of this kind. We will analyze the properties of the matter supporting such spacetime geometries and the linearized  mechanical stability of the shells undergoing perturbations which preserve the cylindrical symmetry. As usual, we adopt units such that $c=G=1$.

\section{Cylindrical shells}

We consider static shells and their perturbative stability analysis. We assume a static background geometry, which allows to define the shells by applying the usual cut and paste procedure starting from two static metrics with cylindrical symmetry.  In cylindrical coordinates  $x^\alpha_{1,2}=(t_{1,2},r_{1,2},\varphi_{1,2},z_{1,2})$ the metrics from which we start can be written in the form
\be \label{le1}
ds^2_{1,2}=-f_{1,2}(r_{1,2})dt^2_{1,2}+f^{-1}_{1,2}(r_{1,2})dr^2_{1,2}+r^2_{1,2}d\varphi^2_{1,2}+\gamma^2_{1,2}r^2_{1,2}dz^2_{1,2}.
\ee
From these geometries we construct a geodesically complete manifold $\mathcal{M}=\mathcal{M}_1 \cup \mathcal{M}_2$, which can take one of the following two forms:
\be \label{enwh}
\mathrm{type\ I:}\;\;\; \mathcal{M}_1=\{x^\alpha_1/0\le r_1\le a_1 \} ,\ \ \ \ \ \mathcal{M}_2=\{x^\alpha_2/ r_2 \ge a_2 \};
\ee
\be \label{wh}
\mathrm{type\ II:}\;\;\;\mathcal{M}_1=\{x^\alpha_1/ r_1 \ge a_1\} ,\ \ \ \ \ \mathcal{M}_2=\{x^\alpha_2/ r_2 \ge a_2 \}.
\ee
with $a_{1,2}$ non null radii. The first case corresponds to joining the interior and the exterior submanifolds while the second case corresponds to joining both exterior regions\footnote{There is a third case corresponding to joining two interior submanifolds, i.e. $\mathcal{M}_1=\{x^\alpha_1/0\le r_1\le a_1 \} ,\  \mathcal{M}_2=\{x^\alpha_2/0\le r_2\le a_2 \}$, that will be not considered in this work.}.   In both cases, the submanifolds are joined at the hypersurface $\Sigma \equiv \partial \mathcal{M}_1 \equiv \partial \mathcal{M}_2 $ defined by
\be
\S: \left\{\begin{array}{ll}H_1(r_1, \t) = r_1 - a_1 (\t) = 0,  \\
H_2(r_2, \t) = r_2 - a_2 (\t) = 0. 
\end{array}\right.
\ee
In order to study the stability under perturbations preserving the symmetry, we have let the radii $a_1(\tau)$ and $a_2(\tau)$ to be functions of the proper time $\tau$ on the shell, which is given by
\be
-d\tau^2=-d{\tau_{1,2}}^2=-f_{1,2}(a_{1,2}){dt_{1,2}}^2+f^{-1}_{1,2}(a_{1,2}){\dot a_{1,2}}^2d\tau^2 ,
\ee
where a dot stands for  a derivative respect to the proper time; thus
\be
d\tau=\frac{f_{1,2}(a_{1,2})}{\sqrt{f_{1,2}(a_{1,2})+{\dot a_{1,2}}^2}}dt_{1,2} .
\ee
By adopting any of the two coordinate systems $\xi_{1,2}^i=(\tau,\varphi_{1,2},z_{1,2})$ on $\S$, the induced metric on the shell is
\be
ds^2=-d\tau^2+a^2_{1,2} d\varphi_{1,2}^2 +  \gamma^2_{1,2}a^2_{1,2}dz_{1,2}^2 .
\ee
Using that $0\le \varphi_{1,2}\le 2\pi $ and the continuity of the geometry across the joining surface (i.e. the continuity of the angular component of the first fundamental form), we find the following relation:
\be \label{fff} 
a_1=a_2=a.
\ee
Then, we can naturally  choose $\varphi = \varphi_{1} = \varphi_{2}$ as the angular coordinate at the matching surface. The coordinates $z_{1,2}$ have to satisfy the condition
\be
\gamma_1dz_1=\gamma_2dz_2
\ee
on the shell. While the geometry must be continuous across the joining surface, the derivatives of the metric are not forced to such restriction. In general, there can be a jump in these derivatives which is associated with the presence of a thin layer of matter. The covariant form of this relation between derivatives of the metric and the matter on the shell are the Lanczos equations \cite{dar,israel}
\be  \label{le2}
8\pi S_{ij}  = [K] h_{ij} - [K_{ij} ],
\ee
where $[K_{ij}] = K_{ij}^{2} - K_{ij}^{1}$ is the discontinuity of the extrinsic curvature tensor across the shell, $[K] = h^{ij} [K_{ij}]$ is the corresponding trace, and $S_{ij}$ is the surface stress-energy tensor. The surface stress-energy tensor satisfies the conservation equation
\be \label{ce}
\nabla _i {S^i}_j = \left[ T_{\alpha \beta} \frac{\partial x^\alpha}{\partial \xi^j}
n^\beta \right],
\ee
where $\nabla _i$ is the covariant derivative on $\Sigma$ and $T_{\alpha \beta}$ is the stress-energy tensor outside  $\Sigma$. The right hand side of Eq. (\ref{ce}) represents a flux from the bulk to the thin shell. The extrinsic curvature tensor is given by
\be
K^{1,2}_{ij} = -n^{1,2}_{\g} \left( \frac{\p^2 x^{\g}_{1,2}}{\p \xi^i \p \xi^j} + (\Gamma_{1,2})^{\g}_{\a \b} \frac{\p x^{\a}_{1,2}}{\p \xi^i} \frac{\p x^{\b}_{1,2}}{\p \xi^j} \right) \Bigg|_{\S}
\ee
and the unit normals to the surface $\S$ pointing from $\mathcal{M}_1$ to $\mathcal{M}_2$ are
\be
n^{1,2}_{\g} = \delta  \Bigg| g^{\a\b}_{1,2} \frac{\p H_{1,2}}{\p x^{\a}_{1,2}} \frac{\p H_{1,2}}{\p x^{\b}_{1,2}} \Bigg|^{-1/2} \frac{\p H_{1,2}}{\p x^{\g}_{1,2}} ,
\ee
with $\delta$ given by
\be
\delta  = \lk \begin{array}{ll}
\;\;+1 , \quad & \mbox{type I geometry},\\
\left.\begin{array}{ll}
-1 ,\; \mathrm{for}\, \mathcal{M}_1 \\
+1 ,\; \mathrm{for}\, \mathcal{M}_2
\end{array} \rk \, & \mbox{type II geometry}.
\end{array}\right.
\ee
The explicit calculation for this general construction yields the normal 4-vector
\be
n^{1,2}_\g= \delta  \left( -{\dot a}, f^{-1}_{1,2}(a) \sqrt{f_{1,2}(a)+{\dot a}^2}, 0, 0\right),
\ee
where we have used that $a_1=a_2=a$ implies $\dot a_1=\dot a_2=\dot a$. In the orthonormal basis on the shell, we have that $h_{\hat i \hat j}=\mathrm{diag}(-1,1,1)$ and the non-vanishing components of the extrinsic curvature tensor
\be
K^{1,2}_{\hat \tau \hat \tau} = - \delta  \frac{2\ddot a + f'_{1,2}(a)}{2 \sqrt{f_{1,2}(a)+{\dot a}^2}},
\ee
\be
K^{1,2}_{\hat \varphi\hat \varphi}= K^{1,2}_{\hat z\hat z} =\delta \frac{ \sqrt{f_{1,2}(a)+{\dot a}^2}}{a}.
\ee
Finally, for the jump $[ K_{\hat{i} \hat{j}} ]$ across the shell we obtain
\be
[K_{\hat \t \hat \t}] = -\frac{2\ddot a+f'_2(a)}{2\sqrt{f_2(a)+{\dot a}^2}}  -\e  \frac{2\ddot a +f'_1(a)}{2\sqrt{f_1(a)+{\dot a}^2}},
\ee
\be
[K_{\hat \varphi\hat \varphi}] = [K_{\hat z\hat z}] =\frac{\sqrt{f_2(a)+{\dot a}^2}}{a} +\e  \frac{\sqrt{f_1(a)+{\dot a}^2}}{a},
\ee
where $\e  =-1$ corresponds to a type I geometry and $\e  =1$  to a type II spacetime. In the orthonormal frame, the surface stress-energy tensor has the form $S_{\hat i \hat j} = \mathrm{diag}(\s, p_\varphi,p_z) $, with $\s$ the surface energy density, and $p_\v$ and $p_z$ the surface pressures. From Eq. (\ref{le2}) we find that the energy density can be written as
\be  \label{sig}
\s  = -\frac{\sqrt{f_2(a)+{\dot a}^2}}{4\pi a} -\e   \frac{\sqrt{f_1(a)+{\dot a}^2}}{4\pi a}.
\ee
Analogously, from Eq. (\ref{le2}), the pressures $p_\varphi= p_z= p$ take the form
\be \label{presphi}
p= \frac{1}{8\pi \sqrt{f_2(a)+{\dot a}^2}}    \lp \ddot a+\frac{f'_2(a)}{2}+\frac{\dot a^2+f_2(a)}{a} \rp + \frac{\e}{8\pi \sqrt{f_1(a)+{\dot a}^2}}    \lp \ddot a+\frac{f'_1(a)}{2}+\frac{\dot a^2+f_1(a)}{a} \rp.
\ee
Clearly the equality of the pressures is not a general feature associated to cylindrical symmetry; rather, this is a consequence of the particular form of the line element, with  metric functions $g_{\varphi\varphi}$ and $g_{zz}$ both proportional to $r^2$. Normal matter satisfies the weak energy condition (WEC), which requires that $\s \ge 0$ and $\s + p \ge 0$. The energy density is always negative for type II geometries, so that the matter on the shell is exotic. For type I geometries, instead, the energy density is positive for $f_2(a_0)<f_1(a_0)$ and is negative for $f_2(a_0)>f_1(a_0)$, so that the corresponding shell can be constituted by normal or exotic matter depending on the values of the parameters. By using Eqs. (\ref{sig}) and (\ref{presphi}), the conservation equation (\ref{ce}) reads
\be \label{cons}
\dot \s + 2\lp\s + p\rp\frac{\dot a}{a}  = 0 ,
\ee
where the vanishing right hand side can be understood as a zero energy flux coming from the bulk; in this case, it is usually said that the shell is transparent.

\section{Stability analysis}

In principle, if the equation of state relating the pressure with the energy density is provided, the equations  (\ref{sig}) and  (\ref{presphi}) above could be integrated to obtain the time evolution of the radius of the shell. However, we are only interested in a slight departure from the radius $a=a_0$ corresponding to a static shell. We then address the stability of a static configuration for which the energy density and pressures take the values $\sigma(a_0)=\sigma_0$ and $p_{\varphi}(a_0)=p_z(a_0)=p_0$, respectively. From the equations above we obtain
\be 
\sigma_0=-\frac{\sqrt{f_2(a_0)}}{4\pi a_0} -\e\frac{\sqrt{f_1(a_0)}}{4\pi a_0}
\ee
and
\be 
p_0=\frac{1}{8\pi}\lp\frac{f'_2(a_0)}{2 \sqrt{f_2(a_0)}}+\frac{\sqrt{f_2(a_0)}}{a_0}\rp +\frac{\e}{8\pi}\lp\frac{f'_1(a_0)}{2 \sqrt{f_1(a_0)}}+\frac{\sqrt{f_1(a_0)}}{a_0}\rp.
\ee
Recalling that $\dot\sigma=\dot a d\sigma/da$, the conservation equation (\ref{cons}) can be recast to
\be \label{consp}
\s' + \frac{2}{a}\lp\s + p\rp  = 0 .
\ee
From Eq. (\ref{sig})  we can write the equation of motion for the shell in the form
\be
\dot{a}^2 + V(a) = 0 ,
\ee 
with the potential
\be
V(a)=\frac{f_1(a)+f_2(a)}{2}-\lp 2\pi a\sigma\rp^2-\frac{\lp f_1(a)-f_2(a)\rp^2}{\lp 8\pi a\sigma\rp^2}.
\ee
A second order Taylor expansion of the potential $V(a)$ around the static solution with radius $a_0$ gives
\be
V(a) = V (a_0) + V'(a_0) (a-a_0) + \frac{1}{2}V''(a_0) (a-a_0)^2 + O(a-a_0)^3.
\ee
It is easy to check that $V(a_0)=0$. The first derivative of the potential, using the conservation equation written in the form  (\ref{consp}), gives $V'(a_0)=0$. 
For a perturbative evolution, i.e., a little departure from the static shell radius, we assume a linearized equation of state for the pressure: 
\be
p = \eta \lp \s-\s_0 \rp + p_0
\ee
where $\eta$ is a constant. Introducing this equation of state and using (\ref{consp}) again, we can calculate the second derivative of the potential evaluated at the equilibrium radius, that is  $V''(a_0)$. The static solution is stable under radial perturbations if $V''(a_0)>0$. This condition solves the problem of the mechanical linearized stability of a cylindrical shell for both type I and type II geometries. 

Introducing the definitions $S(a)=\lp f_1(a)+f_2(a)\rp /2$, $R(a)=\lp f_1(a)-f_2(a)\rp /2$ and $\mu (a)=2\pi a\sigma$, 
the potential can be put in the form
\be
V(a)=S(a)-\mu^2(a)-\frac{R^2(a)}{4\mu^2(a)} ,
\ee
and, after some algebraic manipulations, the second derivative $V''(a_0)$ can be written as
\be
V''(a_0) = \Omega_0 - 2\mu_0\mu_0''- \frac{1}{2\mu_0}\lp\frac{1}{\mu}\rp_0''R_0^2,
 \ee 
where
\be
\Omega_0  =  S_0''-\lb \lp \frac{1}{\mu}\rp _0' \rb^2\frac{R_0^2}{2} -\frac{2}{\mu_0}\lp \frac{1}{\mu}\rp_0 'R_0R_0' - \frac{1}{2\mu_0^2}\lp R_0'^2+R_0R_0''\rp -2 \lp \mu_0 ' \rp ^2 .
\ee
In the expressions above and in what follows, the subscript $0$ stands for functions evaluated at the static radius $a_0$ and for derivatives taken with respect to $a$ and evaluated at $a_0$. From the conservation equation (\ref{consp}) we have $\sigma_0 '=-2(\sigma_0 + p_0)/a_0$; then $\mu_0'=-2\pi(\sigma_0+2p_0)$, so we obtain
\be
\mu_0''=-2\pi \sigma_0' \lp 1+2\eta \rp,
\ee
and
\be
\lp\frac{1}{\mu}\rp_0''=  \frac{2}{\mu_0^3}\lp \mu_0 ' \rp ^2 + \frac{2\pi}{\mu_0^2} \sigma_0' \lp 1+2\eta \rp,
\ee
where $\eta = p_0'/\sigma_0'$. From the definition of $\mu$ we have
\be
\sigma_0'=\frac{1}{2\pi a_0}\lp\mu_0'-\frac{\mu_0}{a_0}\rp,
\ee
so that $V''(a_0)$ can be put in the form
\be
V''(a_0)=  \Omega_0- \chi_0(\eta),
\label{sta}
\ee
where
\be
\chi_0(\eta)= \frac{\lp\mu_0'\rp ^2 R_0^2 }{\mu_0^4} + \frac{2}{a_0}\lp \mu_0 -  \frac{R_0^2}{4\mu_0^3}  \rp \lp \frac{\mu_0}{a_0}-\mu_0'\rp \lp 1+2\eta \rp.
\ee 
As stated before, the condition for stability is that $V''(a_0)>0$. The subsequent analysis of the mechanical stability is carried out in terms of the parameter $\eta $; in the case that $0<\eta\leq 1$ this parameter can be interpreted as the square of the velocity of sound on the shell.

\section{Application to charged shells}

In this section, we present examples of application of the formalism to both type I and II geometries. In all of them, the metrics adopted for the construction of the spacetimes correspond to black string solutions in Einstein--Maxwell theory, given by Eq. (\ref{metric}), with the same negative cosmological constant $\L$ in both regions $\mathcal{M}_1$ and $\mathcal{M}_2$ of the whole manifold $\mathcal{M}$. We start by considering two cases of type I geometries, i.e. we construct spacetimes in which an interior region $\mathcal{M}_1$ is joined by means of a shell $\S $ to an exterior one $\mathcal{M}_2$ that extends to infinity in the radial coordinate.  Then, we analyze two examples of type II spacetimes, corresponding to thin-shell wormholes, with the throat located at the shell $\S $ which joins the two regions $\mathcal{M}_1$ and $\mathcal{M}_2$ that compose $\mathcal{M}$.

\subsection{Shells around vacuum (bubbles)} \label{bubbles}

\begin{figure}[t!] 
\centering
\includegraphics[width=1\textwidth]{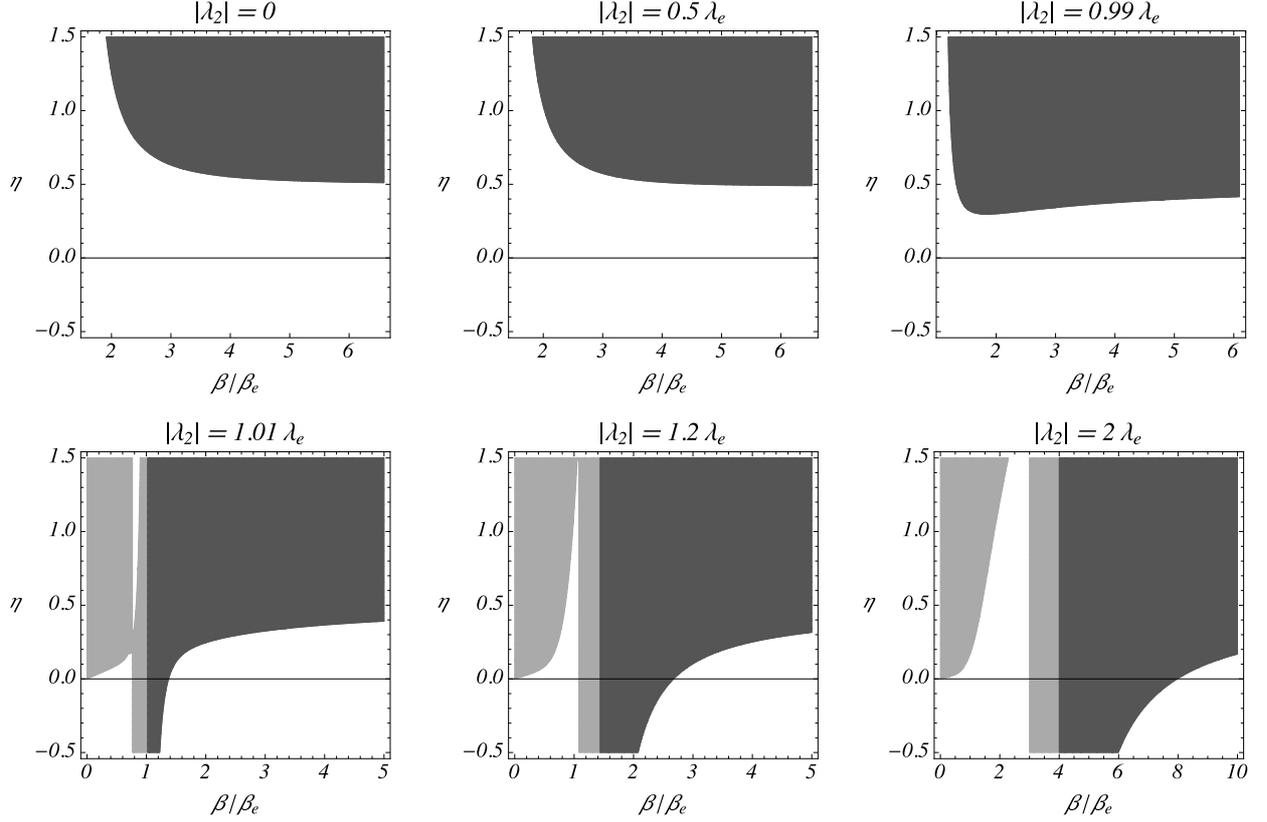}
\caption{Stability regions (light and dark grey) for charged shells with radius $a_0$ around vacuum, in the presence of a negative cosmological constant $\L$. The inner geometry has $m_1=0$ and $\l_1=0$, while the outer one corresponds to a black string characterized by $m_2$ and $\l _2$. Different values of $\l_2/\l_e$ are used in the plots where $\b = a_0 \sqrt{-\L /3}$, $\b_e = m_2^{1/3}$, and $\l_e = m_2^{2/3} \sqrt{3}/2$. The WEC is satisfied in the dark grey region but not in the light grey one.}
\label{f1}
\end{figure}

As a first example, we consider charged shells around vacuum\footnote{Note that in the presence of a non vanishing cosmological constant, the corresponding geometry is not that of the Minkowski spacetime.} (also called bubbles); this is with $m_1 = 0$ and $\l_1 = 0$ in $\mathcal{M}_1$, and a black string geometry with $m_2>0$ and any $\l_2$ in $\mathcal{M}_2$. The interior region has no horizons, so we only need to assume that the radius  $a_0$ of the shell is larger than the event horizon radius of the exterior geometry if $|\l_2| \le \l _e =  m_2^{2/3}\sqrt{3}/2$, in order to avoid the presence of horizons in the whole manifold $\mathcal{M}$. The energy density and the pressure at the shell in this case are given by
\be
\s_0 = \frac{ \a^2 a_0^2- \sqrt{\a^4 a_0^4 - 4  \a a_0 m_2 + 4 \l_2^2 } }{4 \pi \a a_0^2}
\ee
and
\be
p_0 = - \frac{ \a a_0 \sqrt{\a^4 a_0^4 - 4  \a a_0 m_2 + 4 \l_2^2 } + m_2 - \a^3 a_0^3 }{4  \pi a_0 \sqrt{\a^4 a_0^4 - 4  \a a_0 m_2 + 4 \l_2^2 }},
\ee
respectively. The analysis is simplified by defining the dimensionless parameter $\b = a_0 \a $, and the variables
\be \label{uv}
u = \frac{\b}{\b_{e}}  
\qquad \mathrm{and} \qquad
v = \frac{\l_2}{\l_{e}}, 
\ee
with $\b_{e} = m_2^{1/3}$, so the second derivative of the potential takes the form
\be \label{v1}
V'' (a_0) = \frac{ 12 \a^2 \lb u (v^2 - 2 u + 2 u^3 v^2 - u^4) - 
 2 \eta (3 v^4 - 7 u v^2 + 4 u^2 + u^4 v^2 - u^5 )\rb }{u^2 (u^4 -4 u + 3 v^2) (u^2-\sqrt{u^4 - 4 u+ 3 v^2})}.
\ee
Positive values of $V'' (a_0)$ correspond to shells that are stable under radial perturbations. The results are presented in Fig. \ref{f1}, in the plane $u$--$\eta$ for some representative values of $v$. In the upper panel where $v<1$, the range of $u$ is bounded from below by some value that depends on $|\l _2|$, because our construction assumes no horizons in $\mathcal{M}_2$, while in the lower one where $v>1$, no restrictions apply. The regions painted in dark grey correspond to stable configurations where the WEC is fulfilled, i.e. normal matter, whereas the regions in light grey to stable ones but not satisfying this condition, i.e. exotic matter. As normal matter is preferred, we restrict our analysis to this case. We see that there is an important change in the stability behavior near the extremal value $\l _e$ of the modulus of the charge per length  $|\l _2|$. For null or small values of $|\l _2|$, stable solutions require $\eta >0.5$, when $|\l _2|$ gets very close to  $\l _e$ stability is compatible with small and positive values of $\eta $, while for  $|\l _2| > \l _e$ the stable region extends even to negative values of $\eta$. On the other hand, for large values of $u$, in all cases the boundary between the stable and unstable regions asymptotically approaches to $\eta=0.5$.

\subsection{Shells around black strings} \label{shells-bs}

\begin{figure}[t!] 
\centering
\includegraphics[width=1\textwidth]{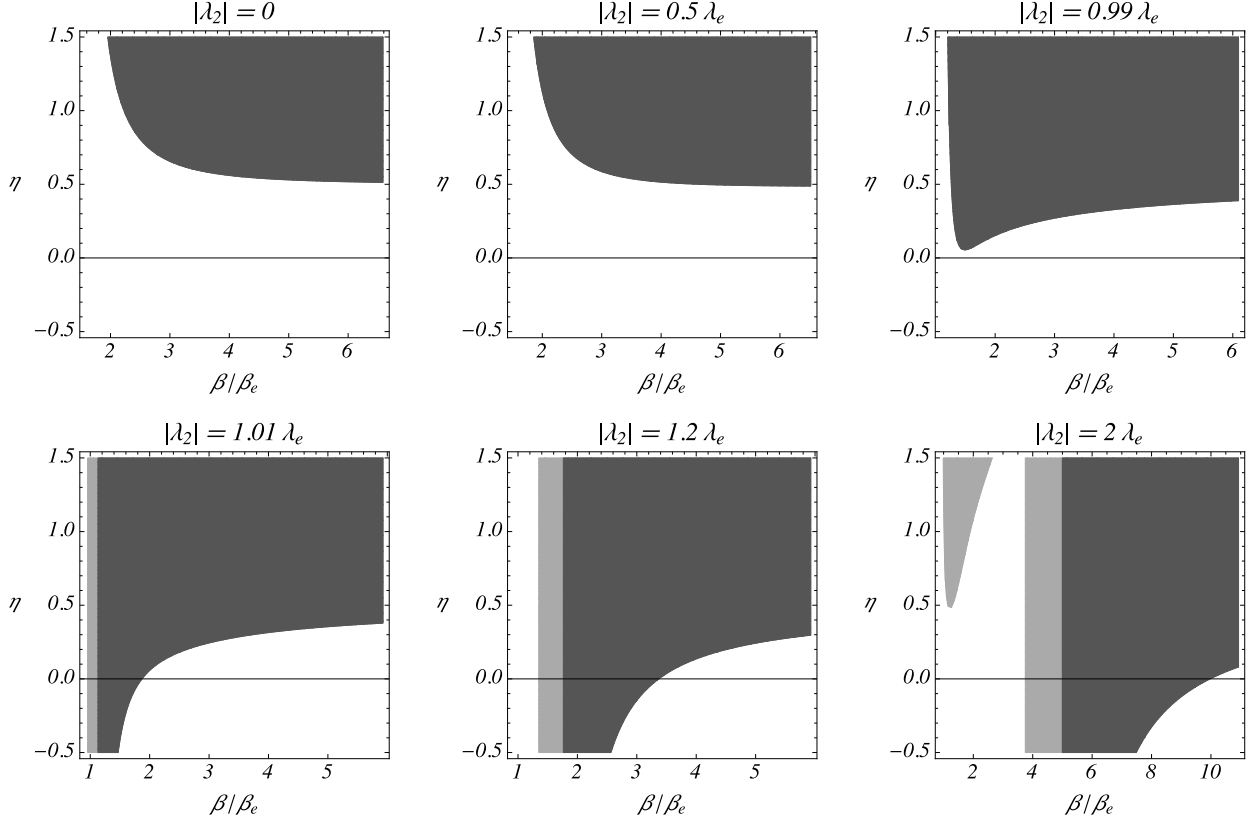}
\caption{Stability regions (light and dark grey) for charged shells with radius $a_0$ around black strings with $m_1 > 0$ and $\l_1=0$, and an exterior geometry with $m_2 > 0$ and any $\l _2$; the negative cosmological constant is $\L $ for the whole spacetime. The relation between the values of the masses per length is $m_1 =0.2 m_2$. Different values of $\l_2/\l_e$ are used in each plot, where $\b = a_0 \sqrt{-\L /3}$, $\b_e = m_2^{1/3}$, and $\l_e = m_2^{2/3} \sqrt{3}/2$. The WEC is satisfied in the dark grey region but not in the light grey one.}
\label{f2}
\end{figure}

\begin{figure}[t!] 
\centering
\includegraphics[width=1\textwidth]{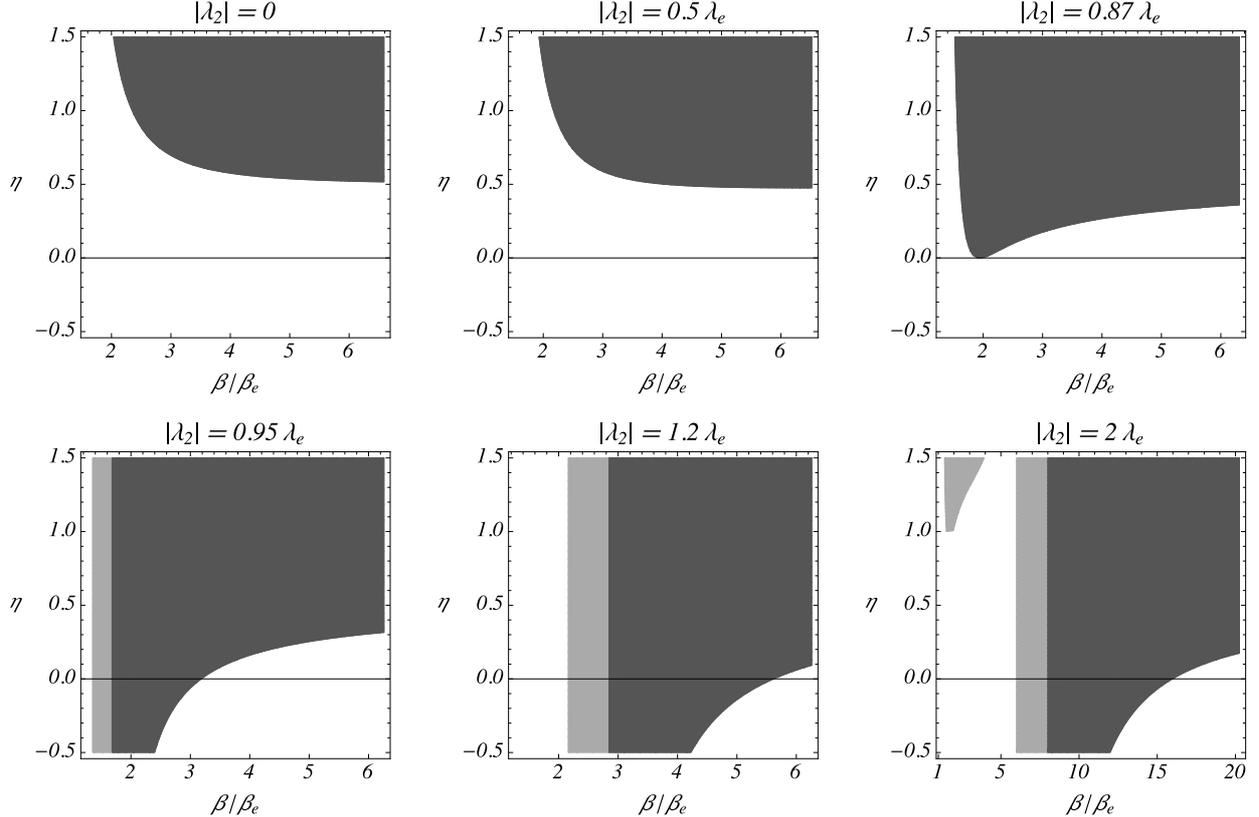}
\caption{Idem Fig. \ref{f2}, now with $m_1 = 0.5 m_2$. In this case, the critical value where the transition occurs is $\l _c = 2^{2/3} 3^{-1/2} \l _e \approx 0.92 \l _e$.}
\label{f3}
\end{figure}

In this second example, we adopt $0 < m_1 < m_2$, $\l_1=0$, and any $\l_2 $. The interior region $\mathcal{M}_1$ corresponds to a black string without charge, with the horizon radius located at $\b _h = \a r_h = (4m_1)^{1/3}$. To have the shell outside the horizon of the inner geometry, we demand that $\b = \a a_0 > \b_h$. The radius  $a_0$ of the shell is also assumed larger than the event horizon radius of the exterior geometry if $|\l _2| \le \l _e =  m_2^{2/3}\sqrt{3}/2$, in order to avoid the presence of horizons in $\mathcal{M}_2$. The energy density and pressure take the form
\be
\s_0 = \frac{ \sqrt{\a ^4 a_0^4 - 4  \a a_0 m_1 } - \sqrt{\a ^4 a_0^4 - 4 \a a_0 m_2 + 4 \l_2^2 } }{4 \pi \a  a_0^2}
\ee
and
\be
p_0 =  \frac{ (m_1- \a^3 a_0^3) \sqrt{\a^4 a_0^4 - 4  \a a_0 m_2 + 4 \l_2^2 } - (m_2 - \a^3 a_0^3 ) \sqrt{\a^4 a_0^4 - 4  \a a_0 m_1  }}{4 \pi a_0 \sqrt{\a^4 a_0^4 - 4  \a a_0 m_1  } \sqrt{\a^4 a_0^4 - 4  \a a_0 m_2 + 4 \l_2^2 }},
\ee
respectively. We adopt again the variables $u$ and $v$ defined in Eq. (\ref{uv}), and 
\be\label{w}
w = \frac{m_1 }{m_2} ,  
\ee
so we can write the second derivative of the potential as
\bea
V'' (a_0) &=&
 \frac{12 \a^2}{(B - A)  (A B)^2 u^4} 
\lk   A^3  u w \lb  2 u w + u^4  \rb + B^3 u \lb   v^2 - 2 u + 2 v^2 u^3 - u^4 \rb
   \right. + \nonumber  \\ 
   &&  \left. 2 \, \eta \, B^2 \lb  B (A^2 u + 4v^2 u  - u^4 v^2 - 3 v^4) -  A^3 u w \rb
   \rk ,
\eea
where 
\be
A = \sqrt{ u^4 -4 u + 3 v^2}, \qquad B = \sqrt{u^4 - 4 u w}. \nonumber
\ee
The stability regions corresponding to $V'' (a_0) > 0$ are shown in Figs. \ref{f2} and \ref{f3} in the plane $u$--$\eta$, for selected values of $v$ and a different value of $w$ in each of them. The regions painted in dark grey represent the stable configurations with normal matter at the shell, while the regions in light grey the stable ones with exotic matter. As before, we are mainly interested in shells with normal matter satisfying the WEC. We can see that:
\begin{itemize}
\item When $\b _h \le \b _e$, or equivalently $0< m_1 \le m_2 /4$, the qualitative behavior of the configurations is very similar to the one found for bubbles in the previous subsection, with the only important difference being that the shell radius should always be larger than the radius of the horizon of the black string, as it can be seen in Fig. \ref{f2} in which $m_1 = 0.2 m_2$. 
\item If $\b _e < \b _h$, or equivalently $m_2 /4 < m_1 < m_2$, the modulus of the charge per length $|\l _2|$ when the stability behavior has a transition moves to $\l _c = (4m_1)^{1/6}(m_2 - m_1)^{1/2}$, which is smaller than the extremal one $\l _e$. This critical value $\l _c$ corresponds to the charge per length for which the horizon $\b _{+} = \a r_{+}$ of the outer geometry used in the construction coincides with the event horizon $\b _h$ of the black string. The value of $\l _c$ becomes smaller in case that $m_1$ approaches to $m_2$. These features are shown in Fig. \ref{f3}, where $m_1 = 0.5 m_2$. 
\end{itemize}
In both cases, the transition happens when the modulus of the charge per length $|\l _2|$ is large enough so that there are no limitations on the radius of the shell $a_0$ due to the outer geometry, so the only restriction on $a_0$ is that it should be larger than the event horizon $r_h$ of the black string.  As in Sec. \ref{bubbles}, for null or small values of $|\l _2|$ the stable solutions occur for $\eta >0.5$; if $|\l _2|$ is close to  $\l _e$ or $\l_c$ depending on the case,  stability is possible for small and positive values of $\eta $, while for  $|\l _2| > \l _e$ or $|\l _2| > \l _c$  as appropriate, the stability zone extends to negative $\eta$. Again, for large $u$, the stable and the unstable regions are asymptotically separated by $\eta=0.5$.

\subsection{Wormholes with the throat joining vacuum and non-vacuum regions} \label{whs1}

\begin{figure}[t!] 
\centering
\includegraphics[width=1\textwidth]{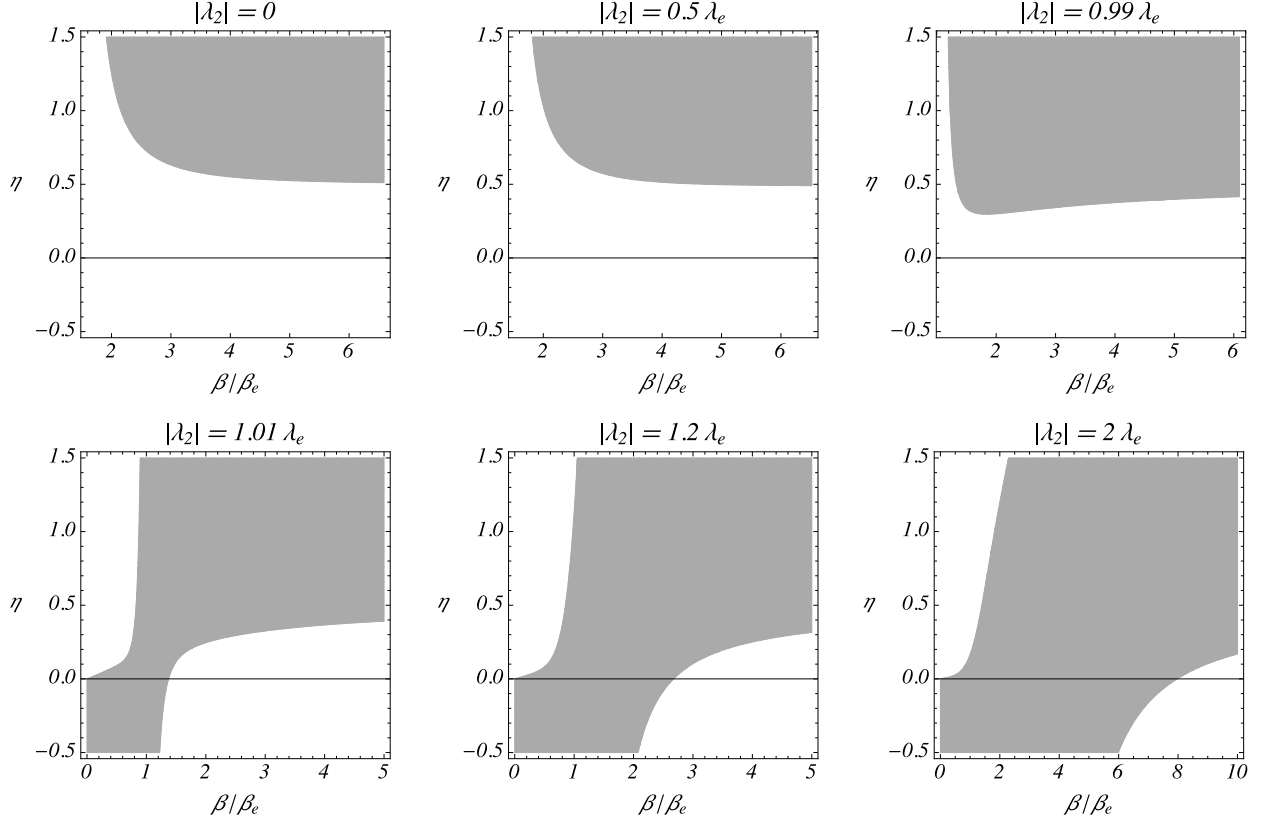}
\caption{Stability regions (light grey) for thin-shell wormholes with a throat radius $a_0$, in two cases of spacetimes with negative cosmological constant $\L$. In the first one, the throat connects a vacuum region, i.e. $m_1=0$ and $\l_1=0$,  with  a black string geometry determined by $m_2$ and $\l_2$. In the second one, two equal black string geometries with $m_1=m_2=m$ and $\l_1=\l_2=\l$ are joined by the throat. In each plot a different value of $\l_2/\l_e$ is used, in all of them $\b = a_0 \sqrt{-\L /3}$, $\b_e = m_2^{1/3}$, and $\l_e = m_2^{2/3} \sqrt{3}/2$. The matter at the throat is always exotic.}
\label{f4}
\end{figure}

We begin by considering wormholes which are not symmetric across the throat. The region $\mathcal{M}_1$ corresponds to a vacuum geometry with $m_1 =0$ and $\l_1 = 0$, and $\mathcal{M}_2$ has the metric of a black string, with $m_2>0$ and any $\l_2$. In order to have a traversable wormhole with no horizons, the radius of the throat $a_0$ should be larger than the event horizon of the metric of the region $\mathcal{M}_2$ when $|\l _2| \le \l _e =  m_2^{2/3}\sqrt{3}/2$. The energy density and pressure are in this case
\be
\s_0 = - \frac{ \a^2 a_0^2  + \sqrt{\a^4 a_0^4 - 4  \a a_0 m_2 + 4 \l_2^2 } }{4 \pi \a  a_0^2}
\ee
\be
p_0 =  \frac{ \a^3 a_0^3 - m_2 + \a a_0 \sqrt{\a^4 a_0^4 - 4  \a a_0 m_2 + 4 \l_2^2 } }{4  \pi  a_0 \sqrt{\a^4 a_0^4 - 4  \a a_0 m_2 + 4 \l_2^2 }},
\ee 
respectively. The negative $\s_0 $ implies that the WEC is not satisfied and the matter is always exotic at the throat. For the stability analysis, we use again the variables $u$ and $v$ introduced in Eq. (\ref{uv}) to write the second derivative of the potential in the form
\be \label{Vwhv}
V'' (a_0) = \frac{ 12 \a^2 \lb u (v^2 - 2 u + 2 u^3 v^2 - u^4) - 
 2 \eta (3 v^4 - 7 u v^2 + 4 u^2 + u^4 v^2 - u^5 ) \rb }{u^2 (u^4 - 4 u + 3 v^2) (\sqrt{u^4 - 4 u+ 3 v^2} + u^2)}.
\ee
The results are graphically presented in Fig. \ref{f4}. In the upper panel, where $v<1$, the values of $u$ are restricted because our construction assumes no horizons in $\mathcal{M}_2$, while in the lower one, where $v>1$, no restrictions apply. The stable solutions correspond to the light grey zones. We appreciate that for null or small values of the charge per length $|\l _2|$, stable solutions are only possible for $\eta >0.5$, when $|\l _2|$ gets close to  $\l _e$ stable shells can be obtained for small and positive values of $\eta $, while for  $|\l _2| > \l _e$ the stable zone is quite large and extends even to negative values of $\eta$. As in the previous examples, for large values of $u$, the boundary between the stable and unstable zones tends to $\eta=0.5$.

\subsection{Wormholes symmetric across the throat} \label{whs2}

This last example corresponds to wormholes which are symmetric across the throat, so we adopt the same metric at both sides of the shell, i.e. we have $m_1=m_2= m > 0$ and any $\l_1=\l_2=\l$. When $|\l | \le \l _e$, the radius $a_0$ is taken larger than the event horizon of the original metric, with the purpose of having a manifold $\mathcal{M}$ without horizons. The energy density and pressure for this case are simply
\be
\s_0 = - \frac{ \sqrt{\a^4 a_0^4 - 4  \a a_0 m + 4 \l^2 } }{2 \pi \a a_0^2}
\ee
and
\be
p_0 =  \frac{ \a^3  a_0^3 - m }{2 \pi a_0 \sqrt{\a^4 a_0^4 - 4  \a a_0 m + 4 \l^2 }},
\ee
respectively. As previously, we use the variables $u$ and $v$ defined in Eq. (\ref{uv}), but now with $\b_e = m^{1/3}$ and $\l_e = m^{2/3}\sqrt{3}/2$, so the second derivative of the potential reads
\be \label{Vwhsim}
V'' (a_0) =  \frac{12 \a^2 \lb u (v^2 - 2 u + 2 u^3 v^2 - u^4) - 
 2 \eta (3 v^4 - 7 u v^2 + 4 u^2 + u^4 v^2 - u^5 ) \rb }{u^4 (u^4 - 4 u + 3 v^2) }.
\ee
The numerators in Eqs. (\ref{Vwhv}) and (\ref{Vwhsim}) are the same, then by considering that both denominators are always positive in these expressions, we obtain in this case the same family of stability regions that in the previous subsection \ref{whs1}, shown in Fig. \ref{f4}, but now with $m_1=m_2=m$ and $\l _1 = \l _2 = \l$. So the same remarks mentioned above apply here.

\section{Summary}

We have presented two classes of cylindrically symmetric thin shells and we have performed the stability analysis of the static configurations under perturbations preserving the symmetry. We have applied the formalism to the study of spacetimes associated to black strings in Einstein--Maxwell theory. In particular, we have mathematically constructed bubbles, thin shells surrounding black strings, and thin-shell wormholes. In the first and second cases, we have found that stable configurations with normal matter are possible if the parameters of the model are suitably chosen. For wormholes, we have obtained that stable static solutions are possible for selected values of the parameters, but they always require exotic matter that violates the weak energy condition. As we have mentioned in the Introduction, the geometry adopted in our construction is closely related to the charged BTZ solution. Then, it is natural to compare our results with those previously  obtained in $2+1$ dimensions \cite{eisi13,eisi14}. For both shells around black holes as for shells supporting wormhole geometries in $2+1$ dimensions, when the modulus of the charge grows, the stability regions enlarge their size and afterwards recover a form similar to that for null or low charges. However, this does not happen in $3+1$ dimensions. On the other hand, the examples studied above of shells associated to black string spacetimes show a sort of improvement in the conditions for stability, when compared with the lower dimensional scenario. For null charge, while the spacetimes in $2+1$ dimensions admitted stable configurations only for values of the parameter $\eta$ larger than unity, now we have found stable configurations compatible with $0\leq \eta<1$. This is a positive feature as, at least for non exotic matter, $\eta$ is usually understood as the squared velocity of sound on the shell, and $\eta>1$ would imply a superluminal wave propagation. Though, in most of the examples studied in this work, the stability with a positive and small $\eta$ requires a large modulus of the charge per length, close to the extremal one. However, we have found the particularly interesting case of charged shells around black strings with $m_2/4 <m_1 <m_2$, in which this transition in the behavior of the stability regions takes place for a modulus of the charge per length parametrically smaller than the extremal one. The effect of perturbations that break the cylindrical symmetry is yet to be investigated, because in some cases this type of perturbations can actually give rise to instabilities (e.g. Gregory--Laflamme instability \cite{grela}).
 
\section*{Acknowledgments}

This work has been supported by Universidad de Buenos Aires and CONICET.

\end{document}